# Topological Theory of Electron-Phonon Interactions in High Temperature Superconductors

J. C. Phillips

Dept. of Physics and Astronomy,

Rutgers University, Piscataway, N. J., 08854-8019

ABSTRACT

There are large isotope effects in the phonon kinks observed in photoemission spectra (ARPES) of optimally doped cuprate high temperature superconductors (HTSC), but they are quite different (Gweon *et al*. 2004) from those expected for a nearly free electron metal with strong electron-phonon interactions (Tang *et al*. 2003). These differences, together with many other anomalies in infrared spectra, seem to suggest that other particles (such as magnons) must be contributing to HTSC. Here we use topological (non-Hamiltonian) methods to discuss the data, emphasizing nanoscale phase separation and the importance of a narrow band of quantum percolative states near the Fermi energy that is spatially pinned to a *self-organized* filamentary dopant array. Topological discrete, noncontinuum, nonperturbative methods have previously explained the form of HTSC phase diagrams without involving detailed microscopic assumptions, and they are especially useful in the presence of strong nanoscale disorder. These methods also explain the "miracle" of an ideal nearly free electron phonon kink in sharply defined nodal quasiparticle states in LSCO at the metal-insulator transition. Finally the universality of the kink energy and Fermi velocity in different cuprates is the result of the marginally elastic nature of these materials, and specifically the isostatic character of the $CuO_2$ planes.

## 1. Introduction

Since the discovery of high temperature superconductivity in the cuprates (Bednorz and Mueller 1986) the foremost question in physics has been what are the interactions responsible for this completely unexpected phenomenon. Mueller himself has always argued that they must be strong electron-phonon interactions, and it was the expectation that these interactions would be strong in generically unstable perovskite-like materials that led to his experiments on doped cuprates. However, the isotope effects on $T_c$ that were the signature of strong electron-phonon interactions in metals have behaved perversely in the ceramic cuprates, being large, as in metals, near the metal-insulator transition and becoming small near optimal doping (Phillips 1991). Indeed the search for isotope effects directly related to superconductivity has seldom been successful (Mueller 2000). This elusive behavior suggested that some other interaction might be involved, but what could this be? Magnons have often been suggested, but these are unreasonable for three major reasons: (1) magnon scattering destroys Cooper pairs, in other words, magnons suppress, not enhance, superconductivity; (2) electron-magnon coupling is much smaller than electron-phonon coupling, and (3) beyond the metal-insulator transition, when $T_c$ is growing towards optimal doping, the filling factor for magnons is decreasing rapidly. Ignoring these reasons, many theorists have continued to use magnons and/or spins to explain various ARPES and infrared anomalies, especially in underdoped samples, but here these anomalies will be explained topologically, with magnons playing only a very minor role.

The disappearance of isotope shifts in $T_c$ near optimal doping could be the result of the optimization process itself (Phillips 1987): small isotope effects "are expected when vibrational frequency changes are compensated by internal coordinate changes". Here we will show that these apparently complex and immeasurable configurational changes are nearly perfectly self-organized (Phillips 2001). This explains quite simply and very accurately the central anomalies in the large isotope shifts recently observed near optimal doping in BSCCO by ARPES (Gweon *et al*. 2004).

**2. The ARPES Kink**

A dispersive kink in $E(\mathbf{k})$ for $\mathbf{k}$ along (11) appears in ARPES spectra for (Pb-doped)BSCCO and LSCO (Lanzara *et al*. 2001). The energy $E_k$ (~ 70 meV) at which the change in slope occurs is somewhat above that of the zone-boundary LO phonon known to be correlated with HTSC (McQueeney *et al*. 1999, 2001), but unambiguously different from magnon energies. The large isotope shifts (Gweon *et al*. 2004) decisively support the phonon interpretation of the dispersive kink, but they raise new problems. In a conventional continnum model (Engelsberg and Schrieffer 1963) dominated by a single Einstein LO phonon mode strongly broadened by many-phonon decay effects (Tang *et al*. 2003), the phonon contribution to the quasiparticle self energy inferred from momentum distribution curves increases linearly with $k - k_F$ up to the phonon cutoff energy $\Theta_D$, and then vanishes. If instead energy distribution curves are used, the phonon contributions below and above $\Theta_D$ are both nearly linear, but of opposite sign. In the absence of many-phonon decay effects the dispersive slope is reduced by a factor $(1 + \lambda)$ up to the phonon cutoff energy $\Theta_D$, and then is unchanged. With many-phonon decay effects in both cases the changes in slope on crossing $\Theta_D$ are reduced by a factor of about 3 for $= \lambda = 1.5$ compared to the unbroadened value of $\lambda$. However, experimentally the isotope shifts for both energy and momentum distributions are *nearly zero* up to the kink energy $E_k$, which is also the phonon cutoff energy $\Theta_D$, and then they begin to increase linearly with $E - E_k$. This nearly perfect cancellation below $\Theta_D$ echoes the cancellation at optimal doping of the isotope shift in $T_c$ itself, and indeed we will show that both effects are nearly equivalent, and are the result of energy level reordering caused by dopant self-organization, an effect not considered in continuum models. Additionally the large isotope effects above $\Theta_D$ continue right up to the point that the signal is incoherently broadened, a seemingly uncontrolled increase that is rarely encountered (a lower cutoff without an upper one).

In addition to this cancellation below $E_k$, the relative isotope shifts change sign as **k** moves from the nodal direction (11) towards the antinodal direction (01). In both directions the slope increases below $\Theta_D$; in the nodal direction (Fig.2a 1) the increase is larger for $^{18}O$ than for $^{16}O$, while the reverse holds as **k** moves closer to the antinodal direction (Fig.2a 6). The increase in slope above $\Theta_D$ means that the effective mass m* has decreased; in nearly free electron models (Lanzara *et al*. 2001) this is interpreted as an undressing of $m^* = m(1 + \lambda)$ below $\Theta_D$ to $m^* = m$ above $\Theta_D$. Ordinarily one would expect a larger effect for heavier masses (normal isotope effect, $\lambda \sim M^{-\alpha}$, with $\alpha \sim 1/2$). However, off-lattice configurational instabilities can produce inverse isotope effects, for instance with interstitial H in Pd (Phillips 1989) because of zero-point motion. In this case one expects the relaxing carriers to move towards $E_F$. Thus the behavior the antinodal direction is "normal", while in the nodal direction it is "inverse", indicative of smaller configurational relaxation.

## 3. Nanoscale Phase Separation and the Participation Ratio

(Pan *et al.* 2001; Lang *et al.* 2002) obtained quantitative pictures of nanodomain structure on very carefully cleaved (*in situ* at low T) micaceous BSCCO by scanning tunneling microscopy (STM). The I-V traces are spatially reproducible on a length scale of 0.15 nm and time scale of weeks, permitting unprecedented surveys of nanodomain and impurity electronic structures. The single most striking aspect of the early data was the near constancy of the nanodomain diameter at 3 nm, independent of doping within the superconductive phase, suggestive of ferroelastic interactions with cutoffs on stress accumulation, consistent with theoretical expectations based on the general properties of perovskites and related ceramics (Phillips 1997, 2003). Subsequent scans have revealed an excellent checkerboard gap structure (Lang *et al.* 2002; McElroy *et al.* 2002) of alternating superconductive (~ 40 meV) and pseudogap (~ 60 meV) 3 nm domains over a field of view of 65 nm$^2$ (~ 40,000 unit cells) near optimal doping. Fourier transforms of the underdoped two-gap structure have shown that there is a charge density wave (CDW) associated entirely with the larger pseudogap (> 60 meV) majority nanodomains, while the lattice structure associated with the smaller superconductive gap pockets (< 60 meV) is undeformed (McElroy *et al.* 2004). This kind of structural pattern is readily explained in terms (for example) of orthorhombic stress patterns that are frustrated and leave behind small residual unstressed pockets near larger nanodomain corners.

Fourier transforms of "octet" subgap states reveal "d-wave symmetry" gap anisotropy proportional to $|\cos 2\phi|$. What is most remarkable about the anisotropy is that in ARPES data and also STM data it holds (McElroy *et al.* 2003) for the superconducting gap within a few percent for optimally and overdoped samples. It is important to note that the d-wave gap projected in this way represents only a small fraction (probably < 10%) of the subgap states, and that the latter are merely a tail on the superconductive density of states itself. Almost all the states responsible for HTSC are not indexable by **k**. The Fourier transforms project gap order on plane waves, much as participation ratios attempt to identify delocalized states in models of metal-insulator transitions. Note that a fraction as small as 10% is inexplicable for a metallic participation ratio, and it cannot be explained merely as a result of fluctuations in two dimensions due to Coulomb interactions (Lee *et al.* 2002), even if the Coulomb interactions are 50 times larger than the band width. In view of the easily deformed nature of perovskites and pseudoperovskites, a much more plausible, material-specific explanation for the observed structure (including charge density waves) is provided by the self-organized internal stress patterns discussed above, and elaborated in considerable detail elsewhere (Phillips *et al.* 2003, Phillips 2004a).

**4. Dopant Self-Organization and Narrow Coherent Percolative Fermi Energy Band**
Conventional percolation theory is inconsistent with the small fraction of percolating gap states. (Phillips 1990, 1999) supposed that there is a narrow energy band of coherent percolative states centered on self-organized dopant configurations (filaments, not stripes, because in the superconductive regions the dopants do not form a superlatttice). The filaments form because with their high conductivities they minimize the free energy by maximizing the enthalpy gained from screening of internal electric fields ("anti-Jahn-Teller effect") (Phillips 1993). The filaments are three-dimensional (Phillips 1990), zigzagging coherently through dopants in semiconductive layers to avoid nanodomain walls in metallic planes: this explains why c-axis conductivities and London carrier densities follow the same scaling relations as those in the ab plane (Homes *et al.* 2004); this scaling relation covers more than six decades and includes Nb and Pb for high carrier densities, where the conductivities are certainly limited by phonon, not magnon, scattering. Lacking such an enthalpy gain, self-organized configurations are exponentially unlikely, much like the exact solution to the traveling salesman problem, said by mathematicians to be "NonPolynomial (NP) complete". Self-organized configurations cannot be identified by algebraic methods applied to continuum Hamiltonians, but their properties are known in several other ways (analogies with network glasses, numerical simulations (Phillips 2002)). Their glassy, strongly disordered nature contrasts with narrow infrared, Raman and neutron line widths in the crystalline superlattice 1/8 "stripe" phase (superconductivity suppressed) of LSCO (Lucarelli *et al.* 2003)); thus these widths are much wider at optimal doping, and percolative filaments should never be confused with crystalline stripes. Such a narrow percolative energy band has often been proposed: a recent example (Ortolani *et al.* 2004) utilized sum rules for the infrared conductivity. The difference $W_n - W_s$ of the integrated planar real part of the conductivity above and below $T_c$ extends to high energies ($\sim$ 3000-4000 cm$^{-1}$) in BSCCO or LSCO because of strong LO phonon interactions at dopants (Hwang *et al.* 2004; Phillips 2004b).

A characteristic feature of this band is its zigzag character: in the ab plane it consists of line segments parallel to Cu-O bonds (Phillips and Jung 2002). The (10) antinodal superconductive states centered on dopant arrays thus have a "strong forward scattering" character when treated by nearly free electron scattering theory (Kulic and Dolgov 2004). Suppose that filamentary formation is enhanced by decreasing the ratio $r = t´/t$ of second neighbor overlap to nearest neighbor overlap in the $CuO_2$ plane, and that such topological enhancement increases $T_c$. (Abrecht *et al.* 2003) observed a reduction in r of a factor of 6 in severely compressed overdoped LSCO films, accompanied by a large decrease in $N(E_F)$. According to continuum models, the decrease in $N(E_F)$ should have reduced $T_c$. In fact, $T_c$ actually increased to a value higher than is found in unstrained optimally doped LSCO, proving that in cuprates topological factors are more important than continuum factors in determining $T_c$.

## 5. Topological Theory

With these elements in place, it is easy to solve the first problem, why the isotope shifts are small both for $T_c$ and for the dispersive E(**k**) *below* the phonon cutoff energy $\Theta_D$. Of course, the attractive well in the gap equation lies below $\Theta_D$ (Choi *et al.* 2003) so the ARPES observation of small isotope shifts below $\Theta_D$ is equivalent to small shifts for $T_c$. But why are the ARPES shifts so small below $\Theta_D$? Constraint theory is a generic theory for dealing with strongly disordered systems (electronic and molecular glasses) (Phillips and Jung 2002). It has explained many of the properties of network glasses (stiffness transitions, reversibility windows (Wang *et al.* 2001)) and the nearly perfect properties of Si/SiO$_2$ interfaces (Lucovsky and Phillips 2004). These ideas also explain the marginal elastic stability of cuprates, which explains why they are the only HTSC (Phillips 2004c). Broadly speaking, constraint theory is a non-Hamiltonian, non-perturbative topological theory that is effective for only a small class of ideally disordered solids, but that class contains exactly those glassy materials, including cuprates, that have posed the hardest problems for conventional algebraic theories.

The central idea of constraint theory is that in a fully self-organized network the strongest constraints will be intact, while the weaker ones are broken (Phillips 1979, 2002). The number of strongest atomic constraints is limited by the discrete number of atomic degrees of freedom of the solid. Here we are concerned with localized LO phonons that bind to carriers associated with dopants to form polarons. All the carriers within $\Theta_D$ of $E_F$ can form polarons by taking advantage of the interaction with an equal number of LO phonons localized near the same number of dopants, with polarization vectors parallel to the local filamentary tangent (the filament need not be linear, and the dopants need not be regularly spaced, as in a commensurate stripe.). The electric dipole oscillator strengths of such coherent filamentary states are far larger than those of any incidental localized states that may coexist elsewhere in the sample, and the excitation of such states dominates optical spectra (infrared or ultraviolet) involving excitation of states below $\Theta_D$. The *integral* number of such LO phonons is assumed to be one/dopant, so the number of constrained carriers matches the number of dopants *exactly*. Of course, this condition is independent of the dopant density (so long as we are in the intermediate phase, which is neither insulating nor normal metallic), in agreement with experiment.

The second problem, the anisotropy of the shifts for dispersive energies more than $\Theta_D$ away from $E_F$, is more difficult. The "normal" isotope shift (states with larger masses shifting more towards $E_F$, that is, relaxation enhances the polarizability) is what we would expect from filamentary states bound to dopants as they relax to improve dielectric screening of the dopant potentials: these states are obviously antinodal states. The nodal states, on the other hand, must be orthogonal to the antinodal filamentary states. (Recall here that ARPES measures only projections of actual disordered states, but this orthogonality is still present after the projection.) Then phase-space incompressibility dictates that as the larger mass antinodal states reorder to move closer to $E_F$, the unbound nodal states should reorder to move away ("inverse" isotope shift). The two shifts approximately cancel, leaving little or no net isotope effect in the angular average even above the phonon cutoff energy $\Theta_D$.

## 6. Absorptive Kink

(Kordyuk *et al*. 2004) have reported a different kind of kink at 100 meV in (Pb-doped)BSCCO, in the *width* of the ARPES peak (related to the quasiparicle scattering rate). The kink appears to shift to ~ 70 meV in optinmally doped BSCCO. This kink in the width of the photoemission peak has a perfectly conventional and straightforward explanation in terms of a distribution of charge density wave (CDW) energy gaps with a maximum cutoff of 100 meV that is temperature-independent up to ~ 150K. According to (McElroy *et al*. 2004), the optimal energy for separating superconductive gaps and CDW gaps spatially is 60 meV, not the ARPES kink value of 70 meV, but the difference between these two values is small compared to their difference with 100 meV. Moreover, because of the projection of spatial inhomogeneities onto **k** involved in ARPES, perfect agreement is not expected. (Kordyuk *et al*. 2004) explained their observed gap in terms of magnon excitations, but the oscillator strength for magnon excitations should be small compared to that for CDW. The shift from 70 meV in BSCCO to 100 meV in (Pb-doped)BSCCO is also more easily explained as a pinning of CDW than as a change in the magnon gap in the $CuO_2$ plane.

## 7. "Normal" LSCO Kink at Metal-Insulator Transition

The primary topic of this paper is the isotope effects (Gweon *et al*. 2004) in nearly optimally doped BSCCO and their explanation in terms of dopant self-organization. The same mechanism also explains the existence of a sharp nodal quasiparticle with a nearly ideal [Fig. 5 of (Tang *et al*. 2003)] dispersive kink in $La_{2-x}Sr_xCuO_4$ at x = 0.063 [Fig.1b of (Zhou *et al*. 2004)], described by the authors as "a miracle". This "miracle" arises in the following way. Filamentary self-organization is in general an exponentially complex matter (Phillips 2004a), as the metallic superconductive filaments must not only avoid each other, but they must also avoid other filaments of different kinds, for instance, charge density waves (McElroy *et al.* 2004), which appear to make the dominant contribution to the pseudogap in underdoped BSCCO. (For example, the self-organized percolation discussed here is much more subtle than random percolation on a lattice, yet fractal discussions of the latter always involve self-avoidance even at threshold, where only one path is involved.). Just above threshold, x = 0.063, the density of metallic self-organized paths is small, but presumably there are still many charge density waves: thus the few metallic self-organized paths must percolate through a forest of spatially inhomogeneous charge density waves. Because the latter are polarized parallel to antinodal (10) directions (Phillips *et al*. 2003), they can do this most easily by going in nodal directions. Well above threshold this would be insufficient, because of collisions with other metallic filaments also percolating in antinodal directions. Thus the conditions for creating an ideal kink are optimal near threshold in LSCO, which is also more stable than BSCCO. Of course, this still does not prove that an ideal kink will be formed, but it does show that the observed ideal kink is far from being accidental. It is also consistent with the general idea that isotope shifts near the metal-insulator transition are "normal", while they decrease as a result of self-organization as optimal doping is approached (Phillips 1987). The self-organization is entirely spatial and orbital, and is unrelated to magnons.

Another feature of the threshold data is that the phonon kink is very narrow (almost a step), in contrast to what one would expect from the conventional continuum model dominated by a single Einstein LO phonon mode strongly broadened by many-phonon decay effects (Tang *et al.* 2003). This requires that many-phonon decay effects disappear near threshold, as happens if the relevant LO phonons are bound to the low density metallic filaments (Phillips 2004a) and actually decay only into similar states on other metallic filaments. [Note that just above the insulator-metal transition at x = 0.06 a zone-boundary LO phonon peak abruptly appears in neutron spectra (McQueeney *et al.* 2001)]. As the doping increases above threshold, the step broadens and at optimal doping only a change in slope is observed (Zhou *et al.* 2003) at $\Theta_D$.

So were the authors (Zhou *et al.* 2004) justified in describing their observation as a "miracle"? Yes, they were, in several ways: a brilliant choice of composition, a perfect cleave of a nonmicaceous, homogeneous sample, ultra high beam intensity to observe a small current, among many experimental factors. But there is also an important theoretical consideration: the ideal kink is exponentially unlikely, in the sense that the cancellation effects observed in nearly optimally doped and much less stable BSCCO (Gweon *et al.* 2004) are absent. The presence of those cancellation effects in the BSCCO optimally doped case is pretty much expectable on the basis of constraint theory, but their disappearance near threshold in LSCO requires a level of sample and surface perfection that can fairly be described as a "miracle".

## 8. Universality of $v_F$ and $\Theta_D$: Another "Miracle"?

The value of the Fermi velocity $v_F$ observed (Zhou *et al.* 2003) in five different cuprates over a wide range of doping *below* $\Theta_D$ is nearly constant, as is $\Theta_D$, much like the nanodomain diameters. The actual phonon cutoffs in the host spectra measured by neutron scattering are not constant, which is, if not a "miracle", at least a major puzzle. Moreover, the slopes (velocities) vary rapidly and monotonically *above* $\Theta_D$, increasing rapidly with decreasing doping right through the metal-insulator transition towards small doping. We again see here an abrupt separation of the observed dispersion curves into two subsets, one constrained to be constant and independent of doping, while the other shows a dramatic chemical trend. Clearly this behavior suggests either intact or broken constraints, with no intermediate cases; such well-defined subsets explain many properties of network glasses (Phillips 1979, 2002). It is just this absence of intermediate cases that distinguishes glasses in their intermediate phases from supercooled liquids.

According to constraint theory (Phillips 2004c), the $CuO_2$ planes are isostatic, while all the other planes are underconstrained (soft and locally buckled). This means that near dopants, located in the planes adjacent to $CuO_2$ planes, or in other metallic planes, k is no longer a good quantum number, and only the phonon and electron states in $CuO_2$ planes show up in ARPES data as resolvable peaks. Moreover, the residual impact of softening associated with unconstrained relaxation in the dopant-deformed underconstrained planes applies only to **v** above $\Theta_D$, accounting for the monotonic decrease of slopes as electron-phonon interactions are screened with increasing dopant concentration. It should be stressed that this Coulombic screening effect is not the result of magnon exchange, as is made abundantly clear by the isotopic dependence of $v_F$ above $\Theta_D$ in the BSCCO experiments (Gweon *et al*. 2004). Note also that that Coulombic screening above $\Theta_D$ has little effect on $T_c$, as the high-energy repulsive interactions above $\Theta_D$ add only a logarithmic correction to the coupling constant $\lambda$, which depends mainly on the strength of the attractive electron-phonon interactions below $\Theta_D$.

## 9. Conclusions

We have shown that discrete "glassy" constraint theory (Phillips 1979, 2002) explains all the major features of the recent ARPES (Gweon *et al*. 2004, Zhou *et al*. 2003, 2004) and STM (McElroy *et al*. 2003, 2004) data. Many strongly correlated continuum "liquid" (polynomial) Hamiltonian models have been proposed to explain various facets of HTSC, involving magnons, CDW's etc., but rarely phonons. The Hamiltonian models met with small successes in explaining individual experiments, but no overall success, and they failed completely to predict large isotope effects. It has even been argued that because isotope shifts on diffraction spectra are small, isotope corrections to superconductivity must be small (Gweon *et al*. 2004). Of course, three-dimensional diffraction spectra cannot probe the effects of glassy (exponentially small fraction of configuration space) self-organization on one-dimensional dopant-centered quantum percolation filaments, which are the central element of the present (and past, going back 15 years!) model.

Constraint theory is an axiomatic, non-Hamiltonian theory that treats self-organized networks in strongly disordered solids; *no new axioms have been introduced here to explain the new ARPES and STM data.* Its successes span a wide range of "insoluble" problems, including even the nature and structure of proteins in their transition states (Rader *et al*. 2003). These widespread successes in describing exponentially unlikely scenarios justify the statement that cuprate HTSC are "almost alive". The self-organized nature of the glassy electronic network explains the various cancellation effects, as well as the seemingly uncontrolled isotope effects above $\Theta_D$.